\documentclass[]{elsart}

\usepackage{amsmath,amsfonts}
\usepackage[dvips]{graphicx}
\usepackage{rotating}
\journal{Physics Letters B}
 
\begin{document}
\begin{frontmatter}
 
\title{Towards observational constraints on a negative 
$(1+z)^4$ type contribution in the Friedmann  equation}
 
\author[WG]{W{\l}odzimierz God{\l}owski}
and
\author[MS,WG]{Marek Szyd{\l}owski}
\address[WG]{Astronomical Observatory, Jagiellonian University,
Orla 171, 30-244 Krak{\'o}w, Poland}
\address[MS]{M. Kac Complex Systems Research Centre, Jagiellonian University,
Reymonta 4, 30-059 Krak{\'o}w, Poland}
 

\begin{abstract}
 We discuss certain issues related to the limitations of density parameters 
for a ``radiation''-like contribution to the Friedmann equation using 
kinematical or geometrical measurements. We analyse the observational 
constraint of a negative $(1+z)^4$-type contribution in cosmological models. 
We argue that it is not possible to determine the energy densities of 
individual components of matter scaling like radiation from astronomical 
observations. We find three different interpretations of the presence of the 
radiation term: 1) the FRW universe filled with a massless scalar field in 
a quantum regime (the Casimir effect), 2) the Friedmann-Robertson-Walker (FRW) 
model in the Randall Sundrum scenario with dark radiation, 3) the cosmological 
model with global rotation. From supernovae type Ia (SNIa) data, 
Fanaroff-Riley type IIb (FRIIb) radio galaxy (RG) data, baryon oscillation 
peak and cosmic microwave background radiation (CMBR) observations we obtain 
bounds for the negative radiation-like term. A small negative contribution 
of dark radiation can reconcile the tension in nucleosynthesis and remove also 
the disagreement between $H_0$ values obtained from both SNIa and the Wilkinson 
Microwave Anisotropy Probe (WMAP) satellite data. 
\end{abstract}
\end{frontmatter}

\section{Introduction} 

In constraining the negative energy contribution to the Friedmann equation, 
a variety of astronomical observations were used, such as SNIa data 
\cite{Riess:2004nr,Astier:2005}, FRIIb RG data \cite{Daly04}, baryon oscillation 
peak and CMBR observations. Although we can obtain the bounds on the sum of  
density parameters for all fluids scaling like radiation, it is impossible to 
determine the contribution coming from each component. In particular it is impossible 
to determine the negative energy density arising from the Casimir effect or dark 
radiation and positive energy density arising from radiation because they scale 
with respect to redshift in the same way in the Friedmann equation. We note that 
the CMBR and big-bang nucleosynthesis (BBN) 
offer stringent conditions on this term, which can be regarded as an established 
upper limit on any individual components of negative energy density, and therefore 
on the Casimir effect, global rotation, discreteness of space following loop 
quantum gravity or brane dark radiation. We argue that even if the precise 
value of the density parameter for fictitious fluid is known from observations, it 
is still not possible to determine the nature (or, say, the origin) of the global 
rotation of the Universe using measurements basing on geometrical optics only, 
i.e. null geodesics $ds^2=0$. For example, if we consider the origin of this 
negative radiation-like term from global rotation effects, then it is not 
possible to come up with different forms of contribution leading to the same 
expression of $-\Omega_{\text{r},0}(1+z)^4$ type. There are, in principle, 
three different interpretations of the presence of the negative 
radiation term in the $H^2(z)$ relation, as follows.

The Casimir effect is a simple observational consequence of the existence
of quantum fluctuation \cite{Bordag01}. The Casimir force between conducting
plates leads to a repulsive force, like the positive cosmological constant. 
Moreover, different laboratory experiments were designed to measure the 
Casimir affect with increased precision and thus strengthen the constraints 
on corrections to Newtonian gravitational law \cite{Fischback98}.
For a survey of recently obtained results in Casimir-energy studies see
\cite{Nesterenko05}. 
For our purpose, it is important that the effect of Casimir energy, which scales 
like radiation, can contribute into the $H^2(z)$ 
relation --- crucial for any kinematical test. It is also interesting that the 
same type of contribution to the effective energy density can be produced by 
loop quantum theory effects in semi-classical quantum cosmology 
\cite{Vandersloot05,Singh05}. These effects give rise to an evolutionary scenario 
in which the initial singularity is replaced by a bounce.
It is worth mentioning that the Casimir-type contributions arising from
tachyon condensation are possible \cite{McInnes06}.

In the brane world scenario, our universe is a submanifold which is embedded 
in a higher-dimensional spacetime, called `bulk space'. In the Randall and 
Sundrum scenario \cite{Randall99}, the Einstein equations restricted to the 
brane reduce to a generalization of the FRW equation with additional terms.
One of these  terms, called dark radiation, is of considerable interest 
because it scales like radiation \cite{Vishwakarma03}. This term arises from 
the non-vanishing electric part of the five dimensional Weyl tensor. Dark 
radiation strongly effects both BBN and CMBR. Ichiki et al. \cite{Ichiki03} 
gave limits on the possible contribution as 
$-1.23<\rho_{dr}/\rho_{\gamma} \le 0.11$ from BBN and 
$-0.41<\rho_{dr}/\rho_{\gamma} \le 0.105$ at the $95\%$ confidence level
from CMBR measurements. Let us note that a small negative contribution of dark 
radiation can also reconcile the tension between the observed 
${}^4\mathrm{He}$ and $D$ abundances \cite{Ichiki02}.

Let us consider Newtonian cosmology following Senovilla et al. 
\cite{Senovilla98}.
Then we can define, following the authors, a homogeneous Newtonian cosmology 
with $\rho$ and $p$ having no spatial dependence, i.e. $\rho=\rho(t)$ and 
$p=p(t)$, and we require that the velocity vector fields depends linearly on 
the spatial coordinates. Then the equation, which represents shear-free 
Newtonian cosmologies with expansion and rotation, well known as the 
Heckmann-Sch{\"u}cking model \cite{Heckmann59} takes the following form
\begin{equation}
\label{eq:5}
\dot{a}^2=\frac{\rho(t_0)}{3a}-\frac{2\omega^2}{3a^2}+C
\end{equation}
where $\omega$ is the rotation scalar and $C$ is an integration constant. 
We interprete it in terms of curvature constant although in the Newtonian 
spacetime the curvature is zero. For our purpose, it is important that 
the effect of rotation produces a negative term in the Newtonian analogue of 
the Friedmann equation. 

In the Newtonian cosmology in contrast to general relativity rotation can 
appear in homogenous and isotropic space. In general relativity the effect 
of rotation are strictly related to non-vanishing shear. The homogeneous 
universe with non-vanishing shear may expand and rotate relative to local 
gyroscopes. The relation between the rotation of the universe and the origin 
of the rotation of galaxies was also investigated 
\cite{Li98,Godlowski03a,Godlowski05,Aryal06}.
Additionally, the role of rotation of objects in the Universe, their 
significance and astronomical measurements was recently addressed by 
\cite{Vishwakarma04,Godlowski03b}).

\section{Observational constraints on the FRW model parameter with a negative radiation-like term}

Usually the fundamental test of a cosmological model is based on the luminosity
distance $d_L$ as a function of redshift \cite{Riess98}.
For the distant SNIa, one can directly observe their apparent magnitude $m$
and redshift $z$. Because the absolute magnitude $\mathcal{M}$ of the
supernovae is related to its absolute luminosity $L$, the relation between 
distance modulus $\mu$,
the luminosity distance $d_L$, the observed magnitude $m$ and the absolute
magnitude $M$ has the following form
\begin{equation} 
\label{eq:11}
\mu \equiv  m - M = 5\log_{10}d_{L} + 25=5\log_{10}D_{L} + \mathcal{M}
\end{equation}
where $D_{L}=H_{0}d_{L}$ and $\mathcal{M} = - 5\log_{10}H_{0} + 25$.
The luminosity distance of a supernova can be obtain as the function of
redshift:
\begin{equation} 
\label{eq:12}
d_L(z) =  (1+z) \frac{c}{H_0} \frac{1}{\sqrt{|\Omega_{k,0}|}} \mathcal{F}
\left( H_0 \sqrt{|\Omega_{k,0}|} \int_0^z \frac{d z'}{H(z')} \right)
\end{equation}
where 
\begin{equation} 
\label{eq:12a}
\left(\frac{H}{H_0}\right)^2= \Omega_{\text{m},0}(1+z)^{3}+\Omega_{k,0}(1+z)^{4}+
\Omega_{\text{r},0}(1+z)^{4}+\Omega_{\text{dr},0}(1+z)^{4}+\Omega_{\Lambda,0},
\end{equation}
$\Omega_{k,0} = - \frac{k}{H_0^2}$ and
$\mathcal{F} (x) \equiv (\sinh (x), x,\sin (x))$ for $k<0, k=0, k>0$,
respectively. We assumed $\Omega_{\text{r},0} = \Omega_{\gamma,0} + \Omega_{\nu,0}
= 2.48 h^{-2} \times 10^{-5} + 1.7 h^{-2} \times 10^{-5}\simeq 0.0001$
\cite{Vishwakarma03}. 
 
Daly and Djorgovski \cite{Daly03} (see also \cite{Zhu04,Puetzfeld05}) 
proposed the inclusion of radio galaxies in the analysis. 
To accomplish this, it is useful to use the coordinate distance 
defined as:
\begin{equation} 
\label{eq:12b}
y(z)=\frac{H_0 d_L(z)}{c(1+z)}.
\end{equation} 

Daly and Djorgovski \cite{Daly04} have compiled a sample comprising the data
on $y(z)$ for 157 SNIa in the Riess et al. \cite{Riess:2004nr} Gold dataset and
20 FRIIb radio galaxies. In our data sets we also include 115 SNIa compiled 
by Astier et~al. \cite{Astier:2005}.

The coordinate distance $y(z)$ does not depend on the value of $H_0$ from definition.
However, when we compute the coordinate distance from the luminosity distance 
(or the distance modulus $\mu$) the knowledge of the value of $H_0$ is required. 
For each sample we choose the values of $H_0$ which were used in original papers. 
We used the distance modulus presented in Ref.~\cite{Riess:2004nr,Astier:2005} for 
the calculation of the coordinate distance. For each sample we choose the values 
of $H_0$ apropriate to the data sets. For Riess et al.'s Gold sample 
we fit the value of $h=0.646$ as the best fitted value and this value is used 
for calculation of coordinate distance for SNIa belonging to this sample.
In turn the value $h=0.70$ was assumed in the calculations
of the coordinate distance for SNIa belonging to Astier et~al.'s sample, 
because the distance moduli $\mu$ presented in Ref.~\cite[Tab.~8]{Astier:2005} 
was calculated with such an arbitrary value of $h=0.70$.
The error of the coordinate distance can be computed as
\begin{equation} 
\label{eq:12c}
\sigma^2(y_i)=\left(\frac{10^{\frac{\mu_i}{5}}}{c \left(1+z \right) 10^5} \right)^2
\left(\sigma^2(H_0)+\left(\frac{H_0 \ln{10}}{5}\right)^2\sigma^2(\mu_i)\right)
\end{equation} 
where $\sigma_i(\mu_i)$ denotes the statistical error of distance modulus 
determination (note that for Astier et~al.'s sample the intrinsic dispersion was also 
included) and $\sigma(H_0)= 0.8$ km/s Mpc denotes error in $H_0$ measurements.

Recently Eisenstein et al. have analysed the baryon oscillation peaks (BOP)
detected in the Sloan Digital Sky Survey (SDSS) Luminosity Red Galaxies 
\cite{Eisenstein:2005}. They found that value of $A$
\begin{equation}
\label{eq:16}
A \equiv \frac{\sqrt{\Omega_{\text{m},0}}}{E(z_1)^{\frac{1}{3}}}
\left(\frac{1}{z_1\sqrt{|\Omega_{k,0}|}}
\mathcal{F} \left( \sqrt{|\Omega_{k,0}|} \int_0^{z_1} \frac{d z}{E(z)} \right)
\right)^{\frac{2}{3}}
\end{equation}
(where $E(z) \equiv H(z)/H_0$ and $z_1=0.35$) is equal $A=0.469 \pm 0.017$.
The quoted uncertainty corresponds to one standard deviation, where a
Gaussian probability distribution has been assumed.
 
It is also possible include in our analysis the so called the (CMBR) ``shift 
parameter''
\begin{equation} 
\label{eq:17}
R \equiv \sqrt{\Omega_{\text{m},0}} \, y(z_{lss})=
\sqrt{\frac{\Omega_{\text{m},0}}{|\Omega_{k,0}|}}
\mathcal{F} \left(\sqrt{|\Omega_{k,0}|} \int_0^{z_{lss}} \frac{d z}{E(z)}\right)
\nonumber
\end{equation}
where $R_0=1.716 \pm 0.062$ \cite{Wang04}.

In our combined analysis, we can obtain a best fit model by minimizing the 
pseudo-$\chi^2$ merit function \cite{Cardone05}:
\begin{equation}
\label{eq:18}
\chi^{2}=\chi_{\text{SN+RG}}^{2}+\chi_{\text{SDSS}}^{2} +\chi_{\text{CMBR}}^{2}=
\nonumber
\end{equation}
\begin{equation}
\sum_{i}\left(\frac{y_{i}^{\text{obs}}-y_{i}^{\text{th}}}{\sigma_{i}(y_{i})}\right)^{2}+
\left(\frac{A^{\text{mod}}-0.469}{0.017}\right)^{2}+
\left(\frac{R^{\text{mod}}-1.716}{0.062}\right)^{2},
\end{equation}
where $ A^{\text{mod}}$ and $R^{\text{mod}}$ denotes value of $A$ and $R$
obtained for particular set of the model parameter.  
For Astier et al. SNIa \cite{Astier:2005} sample additional error in $z$ 
measurements were taken into account. Here $\sigma_{i}(y_{i})$ denotes 
the statistical error (including error in $z$ measurements) of the coordinate 
distance determination.

\begin{figure}
\begin{center}
\includegraphics[width=1.0\textwidth]{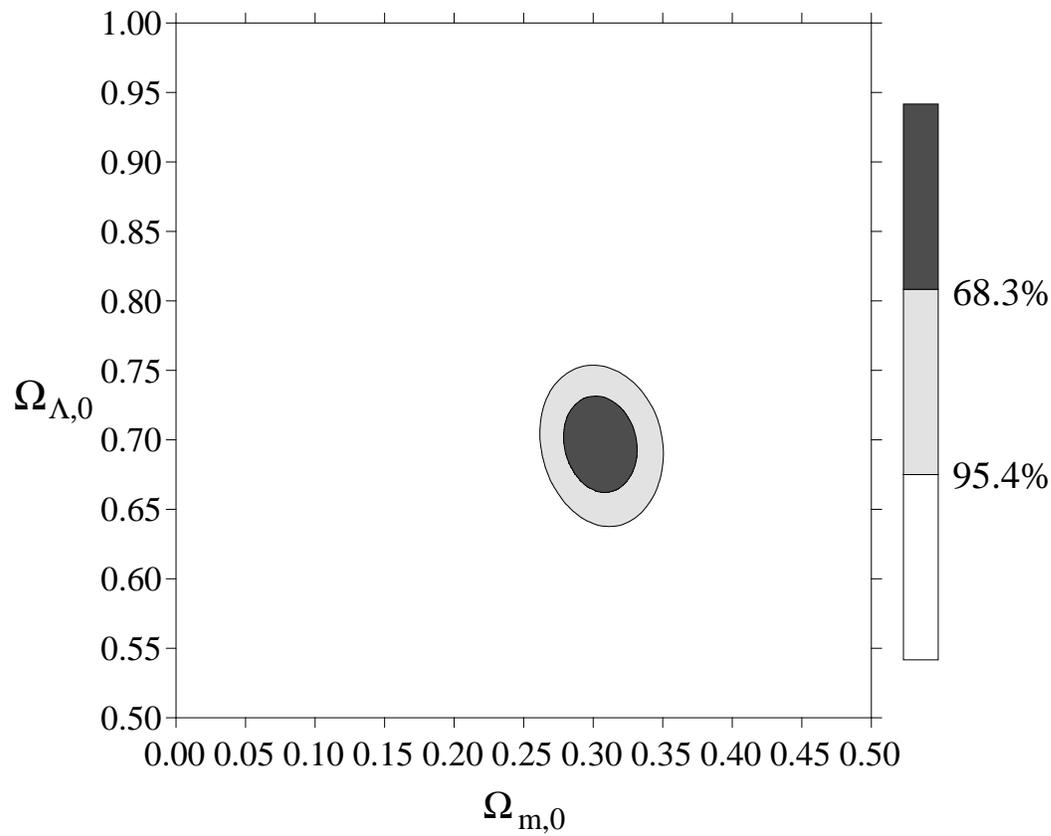}
\end{center}
\caption{The $68.3\%$ and $95.4\%$ confidence levels (obtained from the combined 
analysis of SN+RG+SDSS+CMBR) on the plane ($\Omega_{\text{m},0},\Omega_{\Lambda,0}$).}
\label{fig:1}
\end{figure}

In order to constrain the cosmological parameters, one can minimize
the following likelihood function $\mathcal{L}\propto \exp(-\chi^{2}/2)$.
However, to constrain a particular model parameter, the likelihood function
marginalized over the remaining parameters of the model should be considered
\cite{Cardone05}.
Our results are presented in Table~\ref{tab:1}, Table~\ref{tab:2} and Fig.~\ref{fig:1}.
Table~\ref{tab:1} refers to the minimum $\chi^2$ method, whereas 
Table~\ref{tab:2} shows the results from the marginalized likelihood analysis.

We obtain as the best fit
a flat universe with $\Omega_{\text{m},0}=0.3$, $\Omega_{\text{dr},0}=0$ and
$\Omega_{\Lambda,0}=0.7$. For the dark radiation term, we obtain the stringent 
bound $\Omega_{\text{dr},0}>-0.00035$ at the 95\% confidence level 
($-0.00026$ at the $68.3\%$ confidence level).

Please note that if $\Omega_{\text{r},0}+\Omega_{\text{dr},0}<0$, then we obtain a 
bouncing scenario \cite{Molina99,Tippett04,Szydlowski05} instead of a big bang.
For $\Omega_{\text{m},0}=0.3$, $\Omega_{\text{dr},0}=-0.00035$ and $h=0.65$ 
bounces ($H^2=0$) appear for $z \simeq 1200$. In this case, the BBN epoch never 
occurs and all BBN predictions would be lost. 

\begin{figure}
\begin{center}
\includegraphics[width=0.9\textwidth]{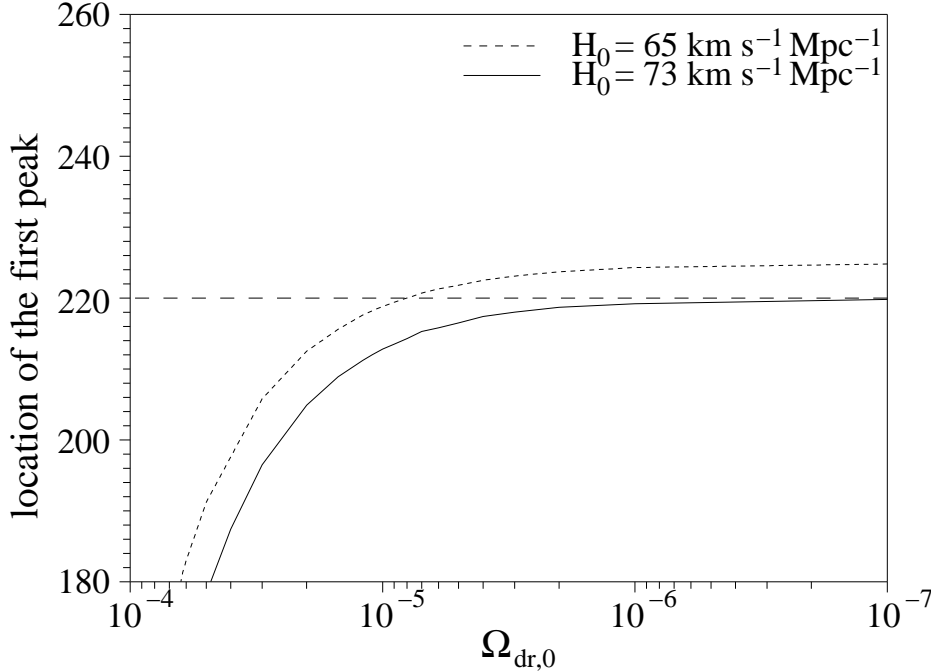}
\end{center}
\caption{The location of the first peak as a function of
$-\Omega_{\text{dr},0}$. Note that $l_1 \simeq 220$ for $h=0.73$ favours 
$\Omega_{\text{dr},0} \simeq 0$, while for $h=0.65$ lead to $\Omega_{\text{dr},0} \ne 0$.}
\label{fig:2}
\end{figure}

To obtain stronger constraints on the model parameters, it is useful to use the 
CMBR observations. The hotter and colder spots in the CMBR can be interpreted 
as acoustic oscillations in the primeval plasma during the last scattering.
It is very interesting that the locations of these peaks are very sensitive to
variations in the model parameters. Therefore, they can be used as another
way of constraining the parameters of cosmological models. The acoustic 
scale $\ell_{A}$ which gives the locations of the peaks is defined as
\begin{equation}
\ell_{A} = \pi \frac{\int_{0}^{z_{\rm dec}} \frac{d z'}{H(z')}}
{\int_{z_{\rm dec}}^{\infty} c_{s} \frac{d z'}{H(z')}}
\end{equation}
where, for the flat model, equation~(\ref{eq:12a}) reduces to
\begin{equation}
H(z) = H_{0} \sqrt{\Omega_{\text{m},0}(1+z)^3 + \Omega_{\text{r},0}(1+z)^4
+\Omega_{\text{dr},0}(1+z)^4  +\Omega_{\Lambda,0}},
\end{equation}
where $c_{\text{s}}$ is the speed of sound in plasma.
Knowing the acoustic scale we can determine the location of the $m$-th peak
$\ell_{m} \sim \ell_{A}(m- \phi_{m})$ where $\phi_{m}$ is the phase shift 
caused by the plasma driving effect. The CMBR temperature angular power 
spectrum provides the locations of the first two peaks 
$\ell_{1} = 220.1_{-0.8}^{+0.8}$,  $\ell_{2} = 546_{-10}^{+10}$ 
\cite{Spergel:2003}. Using three years of WMAP data, Spergel et~al. obtained 
that the Hubble constant $H_{0}=73$ km/s Mpc, the baryonic matter density 
$\Omega_{\text{b},0} = 0.0222h^{-2}$, and the matter density 
$\Omega_{\text{m},0} = 0.128h^{-2}$ \cite{Spergel06}, which are in good agreement with
the observation of position of the first peak (see Fig.~\ref{fig:2}) but lead 
(assuming the $\Lambda$CDM model) to a value $\Omega_{\text{m},0}=0.24$. There 
is also disagreement between $H_0$ values obtained from SNIa and WMAP. We compute 
the location of the first peak as a function of $\Omega_{\text{dr},0}$ assuming 
$H_0=65$ km/s Mpc ($\Omega_{\text{m},0}=0.3$). Now we obtain agreement 
with the observation of the location of the first peak for non-zero values of
the parameter $\Omega_{\text{dr},0}$ (Fig.~\ref{fig:2}). We obtain
$-1.05\times 10^{-5}<\Omega_{\text{dr},0}<-0.5\times 10^{-5}$ at the $95\%$ confidence level. 
Please note that our limit is stronger than that obtained by Ichiki et~al. 
\cite{Ichiki03}, which provides bounds of 
$-7.22 \times 10^{-5}<\Omega_{\text{dr},0} \le 0.65 \times 10^{-5}$ (in the case of the BBN) 
and $-2.41 \times 10^{-5}<\Omega_{\text{dr},0} \le 0.62 \times 10^{-5}$ (in the case of the CMBR). 

All these values of $\Omega_{\text{dr},0}$ are in agreement with the result 
obtained from the combined analysis because the $2\sigma$ confidence interval 
for this parameter obtained from the SNIa data contains the area allowed from 
the CMBR. While the combined analysis gives the possibility that 
$\Omega_{\text{dr},0}$ is equal to zero, the CMBR location of the first peak 
seems to exclude this case for $h=0.65$.

\begin{table}

\caption{Results of the statistical analysis of the model with
the negative radiation term obtained from $\chi^2$ best fit.
The upper section of the table represents the constraint $\Omega_{k,0}=0$ (flat model).}
\begin{tabular}{@{}p{4.2cm}rrrrr}
\hline  \hline
Sample & $\Omega_{k,0}$ & $\Omega_{\text{m},0}$ & $\Omega_{\text{dr},0}$ & $\Omega_{\Lambda,0}$ & $\chi^2$ \\
\hline
SN             &   -   &$ 0.31$ & 0.000 & 0.690 & 297.5     \\
SN+RG          &   -   &$ 0.32$ &-0.001 & 0.681 & 320.4     \\
SN+RG+SDSS     &   -   &$ 0.30$ & 0.000 & 0.700 & 322.4     \\
SN+RG+SDSS+CMBR&   -   &$ 0.30$ & 0.000 & 0.700 & 322.5     \\
\hline
SN             & -0.28 &$ 0.43$ & 0.000 & 0.850 & 296.0     \\
SN+RG          & -0.20 &$ 0.40$ &-0.001 & 0.801 & 319.5     \\
SN+RG+SDSS     &  0.06 &$ 0.29$ & 0.000 & 0.650 & 321.1     \\
SN+RG+SDSS+CMBR&  0.00 &$ 0.30$ & 0.000 & 0.700 & 322.5     \\
\hline
\end{tabular}
\label{tab:1}
\end{table}

\begin{table}

\caption{Results of the statistical analysis of the model with the 
negative radiation term. The values of the model parameters are obtained from 
the marginalized likelihood analysis. We present the maximum likelihood values with 
$68.3\%$ confidence ranges. The upper section of the table represents 
the constraint $\Omega_{k,0}=0$ (flat model).}
\begin{tabular}{@{}p{4.2cm}cccc}
\hline  \hline
Sample & $\Omega_{k,0}$ & $\Omega_{\text{m},0}$ & $\Omega_{\text{dr},0}$ & $\Omega_{\Lambda,0}$ \\
\hline
SN             &   -   & $0.34^{+0.10}_{-0.03}$ & $0.000_{-0.066}$     &$0.66^{+0.02}_{-0.04}$ \\
SN+RG          &   -   & $0.35^{+0.10}_{-0.04}$ & $0.000_{-0.070}$     &$0.65^{+0.02}_{-0.04}$ \\
SN+RG+SDSS     &   -   & $0.31^{+0.02}_{-0.02}$ & $0.000_{-0.007}$     &$0.69^{+0.01}_{-0.01}$ \\
SN+RG+SDSS+CMBR&   -   & $0.30^{+0.02}_{-0.01}$ & $0.00000_{-0.00022}$ &$0.70^{+0.01}_{-0.02}$ \\
\hline
SN             &$-0.44^{+0.24}_{-0.22}$ &$ 0.63^{+0.19}_{-0.15}$ &$0.000_{-0.130}$    &$0.87^{+0.13}_{-0.13}$   \\
SN+RG          &$-0.36^{+0.22}_{-0.22}$ &$ 0.59^{+0.20}_{-0.14}$ &$0.000_{-0.131}$    &$0.82^{+0.12}_{-0.13}$   \\
SN+RG+SDSS     &$ 0.11^{+0.08}_{-0.07}$ &$ 0.29^{+0.02}_{-0.02}$ &$0.000_{-0.023}$    &$0.61^{+0.05}_{-0.06}$   \\
SN+RG+SDSS+CMBR&$ 0.00^{+0.02}_{-0.03}$ &$ 0.30^{+0.02}_{-0.01}$ &$0.00000_{-0.00026}$&$0.70^{+0.02}_{-0.02}$   \\
\hline
\end{tabular}
\label{tab:2}
\end{table}

\section{Conclusion}
 
In our paper we analysed the observational constraints on the negative $(1+z)^4$-type 
contribution in the Friedmann equation. We discussed
different proposals for the presence of such a dark radiation term.
Although it is not possible, with present kinematic observations, to determine 
the energy densities of individual components which scales like radiation, 
we show that some stringent bounds on the value of this total contribution  
can be given. Our detailed conclusions are the following.

\begin{enumerate}

\item The combined analysis of SNIa data and FRIIb radio galaxies using baryon 
oscillation peaks and CMBR ``shift parameter'' give rise to the almost flat universe 
with $\Omega_{\text{m},0} \simeq 0.3$.

\item From the above-mentioned combined analysis, we obtain an upper bound 
$-\Omega_{\text{dr},0}<0.00035$ at the $95\%$ confidence level. This is a stronger limit
than obtained previously by us from SNIa data only \cite{Godlowski03b}.

\item We find new stringent limits on a negative component scaling like radiation
from the location of the peak in the CMBR power spectrum,
$-1.05 \times 10^{-5}<\Omega_{\text{dr},0}<-0.5 \times 10^{-5}$ at the $95\%$ confidence level.
This bound is stronger than that obtained from the BBN and CMBR by Ichiki et~al. \cite{Ichiki03}. 

\item From the limit $-\Omega_{\text{dr},0}<1.05 \times 10^{-5}$ we obtain that
$\Omega_{\text{dr},0}+\Omega_{\text{r},0}>0$. This implies that $H^2(z)$ is always greater
than zero ($H^2(z)>0$) and the bounce does not appear which means that the big bang 
scenario is strongly favoured over the bounce scenario.

\item The discussed model with a small contribution of dark radiation type can also resolve
the disagreement between values of $H_0$ obtained from SNIa and WMAP.

\end{enumerate}

\section{Acknowledgements}

The work of M. S. was supported by project ``COCOS'' No. MTKD-CT-2004-517186 
(during the staying in University of Paris 13). Authors are grateful dr. A. Krawiec 
for fruitful discussion. 
The authors also thank Dr. A.G. Riess, Dr. P. Astier and Dr. R. Daly for the detailed 
explanation of their data samples. We also thanks the anonymous referee for comments
which help us to improve significantly the original version of the letter.

\end{document}